\documentstyle[osa,12pt]{revtex}

\begin{document}
\title{Generation of continuous-wave THz radiation by use of quantum interference}
\author{E.A. Korsunsky and D.V. Kosachiov\thanks{%
permanent adress: Tyumentransgas Co, 627720 Yugorsk, Russia}}
\address{Institut f\"{u}r Experimentalphysik, Technische Universit\"{a}t Graz, A-8010%
\\
Graz, Austria}
\date{\today{}}
\maketitle

\begin{abstract}
We propose a scheme for generation of continuous-wave THz radiation. The
scheme requires a medium where three discrete states in a $\Lambda $
configuration can be selected, with the THz-frequency transition between the
two lower metastable states. We consider the propagation of three-frequency
continuous-wave electromagnetic (e.m.) radiation through a $\Lambda $
medium. Under resonant excitation, the medium absorption can be strongly
reduced due to quantum interference of transitions, while the nonlinear
susceptibility is enhanced. This leads to very efficient energy transfer
between the e.m. waves providing a possibility for THz generation. We
demonstrate that the photon conversion efficiency is approaching unity in
this technique.

OCIS codes: 190.2620, 190.4410, 270.1670, 020.1670

\bigskip
\end{abstract}

Generation of coherent terahertz electromagnetic radiation is a subject of
much current research. This radiation might have interesting potential
applications for electronics, chemical analysis of materials, local radars,
tomography, environment monitoring \cite{THz}, and also for quantum optics
applications and for frequency standards \cite{strumia,god93}. The THz range
is in the gap between the radio-frequency and the visible ranges. Therefore,
neither conventional electronic (microwave) methods nor the photonic ones
can be directly applied to generate coherent THz radiation. Pulsed THz
fields are produced since early seventies and successfully used in different
applications. However, to the best of our knowledge, a reliable source of
{\it continuous-wave} (c.w.) THz radiation is not available up to now.

Here we propose a c.w. generation scheme based on the effect of quantum
interference in multilevel quantum systems (atoms, molecules, dopants in
solids) induced by applied electromagnetic radiation. In an optically dense
media, the interference may be destructive for linear susceptibility
(absorption and refraction), so that an otherwise opaque medium becomes
transparent. This process is termed electromagnetically induced transparency
(EIT) \cite{harr97}. The cancellation of absorption and refraction can also
be explained as being due to the preparation of atoms in a coherent
superposition (''dark'' state) which is immune to applied radiation. When
all the atomic population is perfectly trapped in this superposition, the
medium does not ''see'' the radiation. It turns out, however, that if the
dark state is slightly disturbed the linear susceptibility remains still
very small, while the nonlinear susceptibility is resonantly enhanced by
constructive interference \cite{harr90}. Therefore, nonlinear frequency
conversion/generation processes are very efficient in such media. Several
successful experiments have been performed demonstrating efficient
generation of XUV radiation with atomic hydrogen \cite{stoi}, red to blue
frequency conversion with molecular sodium \cite{welleg}, enhanced four-wave
mixing with doped crystals \cite{hemm97}. Harris with coworkers have reached
blue to UV \cite{jain96} and UV to VUV \cite{mer99} conversion in atomic Pb
vapor with almost unity photon-conversion efficiency. The aim of the present
paper is to extend these ideas to the case of THz generation, treating an
interaction of the e.m. radiation with atoms as well as the propagation of
radiation through the medium in exact manner.

We suggest that the THz radiation can be produced by the EIT-assisted
frequency conversion in media where three discrete states may be selected,
two of them being metastable ones. An example of such a scheme is a closed
three-level $\Lambda $ system (Fig. 1), where $\left| 1\right\rangle -\left|
3\right\rangle $ and $\left| 2\right\rangle -\left| 3\right\rangle $ are the
dipole-allowed optical transitions, and the frequency of a magnetic-dipole
transition $\left| 1\right\rangle -\left| 2\right\rangle $ is in the THz
range. Such systems can be found in vapors of alkaline-earth atoms with the
corresponding states $\left| 1\right\rangle \equiv \,^3P_0$ or $\,^3P_2$, $%
\left| 2\right\rangle \equiv \,^3P_1$ and $\left| 3\right\rangle \equiv
\,^3S_1$. Frequencies of the transitions $\,^3P_0\,-$ $\,^3P_1$ and $%
\,\,^3P_2\,-\,^3P_1$ are 0.6 and 1.2 THz for $^{24}$Mg, 1.5 and 3.2 THz for $%
^{40}$Ca, etc. A similar $\Lambda $ system can be excited, for instance, in
samarium atoms where the transition between two substates $\,4f^66s^2\,^7F_0$
and $^7F_1$ of the ground state fine structure has a frequency of 8.78 THz.
One may also find appropriate schemes in other atoms, where the THz
transition is usually between the two fine structure components, and
probably in some molecules and doped crystals as well.

Let us assume that the $\Lambda $ atom interacts with a bichromatic optical
field
\begin{equation}
{\bf E}(z,t)=\,\sum\limits_{m=1,2}{\bf e}_{3m}E_{3m}(z,t)\frac 12\exp \left[
-i\left( \omega _{3m}t-k_{3m}z+\varphi _{3m}(z,t)\right) \right] +c.c.
\label{E}
\end{equation}
having the frequencies $\omega _{3m}$ resonant or near-resonant with
transitions $\left| m\right\rangle -\left| 3\right\rangle ,\,\,m=1,2$, and
with the radiation of THz frequency $\omega _T$%
\begin{equation}
{\bf H}(z,t)=\,{\bf e}_TH(z,t)\frac 12\exp \left[ -i\left( \omega
_Tt-k_Tz+\varphi _T(z,t)\right) \right] +c.c.,  \label{H}
\end{equation}
resonant with transition $\left| 1\right\rangle -\left| 2\right\rangle $.
Here, $E_{3m}$ and $H$ are the amplitudes, ${\bf e}_{3m}$ and ${\bf e}_T$
are the unit polarization vectors, $\varphi _{3m}$ and $\varphi _T$ are the
phases, and $k_{3m}=\omega _{3m}/c$ and $k_T=\omega _T/c$ are the
wavenumbers of electric and magnetic field, respectively. Both the
amplitudes $E_{3m}$,$\,\,H$ and the phases $\varphi _{3m}$, $\varphi _T$ are
regarded as slowly varying functions of time and coordinate.

The propagation of e.m. waves along the $z$ axis in the medium is governed
by the Maxwell equations. In the slowly varying amplitude and phase
approximation, these equations can be transformed to the following form \cite
{scbook,kos95,kors99}:
\begin{mathletters}
\label{Maxw1}
\begin{eqnarray}
\frac{\partial E_{3m}}{\partial z}+\frac 1c\frac{\partial E_{3m}}{\partial t}
&=&-N\frac{4\pi d_{3m}\omega _{3m}}c%
\mathop{\rm Im}%
(\tilde{\sigma}_{3m}), \\
\frac{\partial \varphi _{3m}}{\partial z}+\frac 1c\frac{\partial \varphi
_{3m}}{\partial t} &=&-N\frac{4\pi d_{3m}\omega _{3m}}c\frac 1{E_{3m}}%
\mathop{\rm Re}%
(\tilde{\sigma}_{3m}), \\
\frac{\partial H}{\partial z}+\frac 1c\frac{\partial H}{\partial t} &=&-N%
\frac{4\pi \mu \omega _T}c%
\mathop{\rm Im}%
(\tilde{\sigma}_{21}), \\
\frac{\partial \varphi _T}{\partial z}+\frac 1c\frac{\partial \varphi _T}{%
\partial t} &=&-N\frac{4\pi \mu \omega _T}c\frac 1H%
\mathop{\rm Re}%
(\tilde{\sigma}_{21}),
\end{eqnarray}
where $d_{3m}\equiv \left| {\bf e}_{3m}{\bf d}_{3m}\right| ,\,\mu \equiv
\left| {\bf e}_T\vec{\mu}\right| $, ${\bf d}_{3m}$ and $\vec{\mu}$ are the
matrix elements of the electric-dipole moment operator ${\bf \hat{d}}$ and
the magnetic-dipole moment operator ${\bf \hat{\mu}}$, respectively, in the
basis of bare atomic states $|n\rangle ,\,n=1,2,3$: ${\bf d}_{3m}\equiv
\langle 3|{\bf \hat{d}}|m\rangle $, $\vec{\mu}\equiv \langle 1|{\bf \hat{\mu}%
}|2\rangle $; $N$ is the density of the active atoms. The medium
polarization components (the right-hand side of Eqs. (\ref{Maxw1})) are
determined by the density matrix elements $\sigma _{ns}$ averaged over the
atomic velocities with the distribution $w\left( v_z\right) $ where $v_z$ is
the $z$-projection of the atom velocity:
\end{mathletters}
\begin{equation}
\tilde{\sigma}_{ns}=\int\limits_{-\infty }^{+\infty }dv_z\,w\left(
v_z\right) \sigma _{ns},  \label{sigma-av}
\end{equation}
with
\begin{equation}
\sigma _{ns}=\rho _{ns}\left( v_z\right) \exp \left[ i\left( \omega
_{ns}t-k_{ns}z+\chi _{ns}\right) \right] ,  \label{sigma}
\end{equation}
where $\rho _{ns}\equiv \langle n|\hat{\rho}|s\rangle $, ${\hat{\rho}}$ is
the atomic density matrix; and the phase $\chi _{ns}$ is the sum of the e.m.
field phase $\varphi _{ns}$ and the phase $\vartheta _{ns}$ of the atomic
dipole moment ${\bf d}_{3m}=\left| {\bf d}_{3m}\right| {\rm e}^{i\vartheta
_{3m}}$, ${\bf \vec{\mu}}=\left| {\bf \vec{\mu}}\right| {\rm e}^{i\vartheta
_{12}}$: $\chi _{3m}=\varphi _{3m}+\vartheta _{3m}$, $\chi _{12}=\varphi
_T+\vartheta _{12}$.

The $\Lambda $ system represents a typical interaction scheme where the
interference of excitation channels may lead to creation of a dark state and
preparation of atoms in this state \cite{cptrev}. If the e.m. field between
the states $\left| 1\right\rangle $ and $\left| 2\right\rangle $ is not
applied then the following superposition of the metastable states
\begin{equation}
\left| NC\right\rangle =\frac{g_{32}/g_{31}}{\sqrt{1+g_{32}^2/g_{31}^2}}%
\left| 1\right\rangle -\exp \left( \chi _{32}-\chi _{31}\right) \frac 1{%
\sqrt{1+g_{32}^2/g_{31}^2}}\left| 2\right\rangle ,  \label{NC}
\end{equation}
gets completely decoupled from the field when the two-photon resonance
condition is satisfied:
\begin{equation}
\Delta _{32}-\Delta _{31}=0.  \label{2ph-cond}
\end{equation}
In Eqs. (\ref{NC}) and (\ref{2ph-cond}), $g_{3m}=d_{3m}E_{3m}/2\hbar $ are
the Rabi frequencies of the optical fields, $\Delta _{3m}=\omega
_{3m}-\left( {\cal E}_3-{\cal E}_m\right) /\hbar $ are the laser frequency
detunings from transitions $\left| m\right\rangle -\left| 3\right\rangle
,\,(\,m=1,2)$, ${\cal E}_n$ is the eigenenergy of the atomic state $\left|
n\right\rangle $. The superposition $\left| NC\right\rangle $ is stable, and
it is fed by spontaneous emission from the level $\left| 3\right\rangle $ so
that all the atomic population is trapped in $\left| NC\right\rangle $ after
some optical pumping time. Therefore, the light ceases to interact with
atoms and propagates through the medium without absorption and refraction.
We note that this occurs for any relation between the laser intensities and
phases.

At the same time, the preparation of atoms in the superposition of
metastable states $\left| 1\right\rangle $ and $\left| 2\right\rangle $
means that the coherence $\sigma _{12}$ is induced. Corresponding to the
Eqs. (\ref{Maxw1} c,d), this should lead to a generation of the field with
frequency $\left( \omega _{31}-\omega _{32}\right) $. Such an EIT-assisted
generation of a microwave radiation has recently been observed in atomic Cs
vapor \cite{god98}. In general, the presence of a third e.m. field destroys
the dark state, so that the optical waves start to interact with the medium.
However, the degree of the destruction is very small if the generated field
is weak. In the present case, the THz generation corresponds to the
difference-frequency generation scheme \cite{boyd}. If we suppose the photon
conversion efficiency to be unity, that is for each photon of frequency $%
\omega _{31}$ one THz-photon (and one photon of frequency $\omega _{32}$) is
generated, then the maximum energy (or power) conversion efficiency is given
by $\eta =\hbar \omega _T/\hbar \omega _{31}\ll 1$. For example, for the $%
\Lambda $ system ($3^3P_1-3\,^3P_2-$ $\,4^3S_1$) in $^{24}$Mg, we have $%
\omega _T=1.22\cdot 10^{12}$ Hz, $\omega _{31}=5.80\cdot 10^{14}$ Hz, and $%
\eta =0.0021$. Therefore, the THz intensity is always much smaller than the
optical one, so that the dark state disturbance is always very small. Thus,
the total e.m. energy dissipation in the medium is negligible, and one can
hope to get a very efficient generation in such a nearly dark (''grey'')
medium.

In this paper we concentrate on a regime of the radiation propagation in a
continuous wave limit, that is, we shall assume that the characteristic time
of a change in the field amplitudes and phases, and the interaction time of
atoms with the radiation are much longer than the characteristic time of a
change in the internal state of an atom. Then, the time derivatives in Eqs. (%
\ref{Maxw1}) can be dropped, and the steady-state values of the density
matrix elements $\sigma _{ns}$ can be used in the right-hand side of Eqs. (%
\ref{Maxw1}). The system considered here is an example of the so-called
closed-loop system, where the steady state can be established only when the
multi-photon resonance condition is satisfied \cite{kos92}:
\begin{equation}
\omega _{31}-\omega _{32}-\omega _T=0.  \label{mp-res}
\end{equation}
In what follows we will always assume the condition (\ref{mp-res}) to be
fulfilled.

The density matrix equations for the $\Lambda $ atom can be found in Refs.
\cite{kos95,kos92}. These equations can be solved analytically under some
simplifying conditions. We suppose that spontaneous relaxation rates for the
channels $\left| 3\right\rangle \rightarrow \left| 1\right\rangle $ and $%
\left| 3\right\rangle \rightarrow \left| 2\right\rangle $ are equal: $\gamma
_{31}=$ $\gamma _{32}\equiv \gamma $. With the condition $\omega
_{31}\approx \omega _{32}$ (since $\left( \omega _{31}-\omega _{32}\right)
/\omega _{31}=\omega _T/\omega _{31}\ll 1$) this would also imply $\left|
{\bf d}_{31}\right| \approx \left| {\bf d}_{32}\right| $. This
simplification allows us to get analytical results which demonstrate all the
basic features of the process. Numerical calculations with real atomic
parameters (Figs. 2-4) confirm the analytical results not only
qualitatively, but even quantitatively. This suggests that the performance
of the proposed scheme is not very sensitive to exact values of the
relaxation rates and dipole moments on optical transitions. Another
assumption used is quite reasonable. Since the THz radiation intensity is
very small, we assume that its Rabi frequency is small: $g_T=\mu H/2\hbar $ $%
\ll \gamma $.

Solution of the steady-state density matrix equations for $v_z=0$, $\Gamma
=0 $ ($\Gamma $ is the relaxation rate of the coherence between states $%
\left| 1\right\rangle $ and $\left| 2\right\rangle $), equal spontaneous
relaxation rates: $\gamma _{31}=\gamma _{32}\equiv \gamma $, and zero
detunings $\Delta _{31}=\Delta _{32}=0$, gives to the first order in $\left(
g_T\,/\,\gamma \right) $:
\begin{mathletters}
\label{ims}
\begin{eqnarray}
\mathop{\rm Im}%
(\sigma _{31}) &=&\frac{g_Tg_{32}}{g_0^2}\sin \Phi , \\
\mathop{\rm Im}%
(\sigma _{32}) &=&-\frac{g_Tg_{31}}{g_0^2}\sin \Phi , \\
\mathop{\rm Im}%
(\sigma _{21}) &=&-\frac{g_{31}g_{32}}{g_0^2}\sin \Phi ,
\end{eqnarray}
\end{mathletters}
\begin{mathletters}
\label{res}
\begin{eqnarray}
\mathop{\rm Re}%
(\sigma _{31}) &=&-\frac{g_Tg_{32}\left( g_{32}^2-g_{31}^2\right) }{g_0^4}%
\cos \Phi , \\
\mathop{\rm Re}%
(\sigma _{32}) &=&\frac{g_Tg_{31}\left( g_{32}^2-g_{31}^2\right) }{g_0^4}%
\cos \Phi , \\
\mathop{\rm Re}%
(\sigma _{21}) &=&-\frac{g_{31}g_{32}}{g_0^2}\cos \Phi ,
\end{eqnarray}
The populations of the metastable states are
\end{mathletters}
\begin{eqnarray*}
\rho _{11} &=&\frac{g_{32}^2}{g_0^2}-\frac{2g_Tg_{31}g_{32}\gamma }{g_0^4}%
\sin \Phi , \\
\rho _{22} &=&\frac{g_{31}^2}{g_0^2}+\frac{2g_Tg_{31}g_{32}\gamma }{g_0^4}%
\sin \Phi ,
\end{eqnarray*}
where $g_0^2=g_{31}^2+g_{32}^2$, and the relative phase $\Phi $ is
determined as
\begin{equation}
\Phi =\left( \chi _{31}-\chi _{32}\right) -\chi _{12}.  \label{Phi}
\end{equation}
The excited state population $\rho _{33}$, which is responsible for
irreversible dissipation of the e.m. energy by the medium, is of the second
order in $\left( g_T\,/\,\gamma \right) $. This indicates that atoms are
really in the grey state. One sees from Eqs. (\ref{ims}) and (\ref{res}),
that the medium is absolutely transparent and not refractive for
\begin{equation}
\Phi =\pi n,\,\,n=0,1,2,...  \label{phi-cond}
\end{equation}
and
\begin{equation}
g_{31}=g_{32}.  \label{g-cond}
\end{equation}
These are exactly the conditions for the dark state in closed $\Lambda $
system \cite{kos95,kos92,buck86,kos91}.

For arbitrary optical field amplitudes and phases, however, the refraction
and absorption (or amplification) of individual frequency components may be
substantial. The change of the fields can be calculated analytically by the
method developed in paper by Armstrong {\it et. al.} \cite{arm62}. We insert
the density matrix elements (\ref{ims}) and (\ref{res}) in the Maxwell
equations (\ref{Maxw1}). The amplitude equations are read then:
\begin{mathletters}
\label{Maxw2}
\begin{eqnarray}
\frac{dE_{31}}{dz} &=&-\frac{\pi N}{\hbar ^2c}\frac{d_{31}d_{32}\mu }{g_0^2}%
\omega _{31}E_{32}H\sin \Phi , \\
\frac{dE_{32}}{dz} &=&\frac{\pi N}{\hbar ^2c}\frac{d_{31}d_{32}\mu }{g_0^2}%
\omega _{32}E_{31}H\sin \Phi , \\
\frac{dH}{dz} &=&\frac{\pi N}{\hbar ^2c}\frac{d_{31}d_{32}\mu }{g_0^2}\omega
_TE_{31}E_{32}\sin \Phi .
\end{eqnarray}
The intensities associated with each of these waves are given by $I_{3m}=$ $%
\left( c/8\pi \right) E_{3m}^2\,$and $\,I_T=$ $\left( c/8\pi \right) H^2$.
The set of equations (\ref{Maxw2}) shows that the total power flow
(proportional to the total intensity $I=I_{31}+I_{32}+I_T$) is conserved, as
expected for propagation through a lossless (due to EIT) medium:
\end{mathletters}
\[
\frac{dI}{dz}=\frac{dI_{31}}{dz}+\frac{dI_{32}}{dz}+\frac{dI_T}{dz}=\frac N{%
4\hbar ^2}\frac{d_{31}d_{32}\mu }{g_0^2}\left( \omega _{32}+\omega _T-\omega
_{31}\right) E_{31}E_{32}H\sin \Phi =0,
\]
where the last equality follows from the multiphoton resonance condition (%
\ref{mp-res}). Obviously, the energy losses (proportional to the population
of the excited state $\rho _{33}$) appear only in the second order in $%
\left( g_T\,/\,\gamma \right) $.

The set of equations (\ref{Maxw2}) also implies that
\begin{equation}
\frac d{dz}\left( \frac{I_{31}}{\hbar \omega _{31}}\right) =-\frac d{dz}%
\left( \frac{I_{32}}{\hbar \omega _{32}}\right) =-\frac d{dz}\left( \frac{I_T%
}{\hbar \omega _T}\right) ,  \label{M-R}
\end{equation}
which are the well-known Manley-Rowe relations \cite{boyd}. These relations
tell us that the rate at which photons at frequency $\omega _T$ are created
is equal to the rate at which photons at frequency $\omega _{32}$ are
created and is equal to the rate at which photons at frequency $\omega _{32}$
are destroyed.

In order to solve the Maxwell equations, we introduce the following
dimensionless variables: field amplitudes
\begin{eqnarray*}
u_m &=&\left( \frac{8\pi }cI\frac{\omega _{3m}}{\omega _0}\right)
^{-1/2}E_{3m}\,,\,\,\,(m=1,2), \\
u_T &=&\left( \frac{8\pi }cI\frac{\omega _T}{\omega _0}\right) ^{-1/2}H,
\end{eqnarray*}
where $\omega _0=\left( \omega _{31}+\omega _{32}+\omega _T\right) /2$, and
a dimensionless optical length $\zeta =\kappa z$ with the coefficient
\[
\kappa =\frac{4\pi N}c\frac{d_{31}d_{32}\mu }{g_0^2}\left( \frac{8\pi }c%
\frac I{\omega _0}\right) ^{1/2}\sqrt{\omega _T\omega _{31}\omega _{32}}.
\]

As we have discussed above, the THz intensity is always much smaller than
the optical one: $I_T\ll I_{31}+I_{32}$. Therefore, we can safely assume
that the total optical intensity $I_{31}+I_{32}\approx I$ is conserved. This
follows also from the first of the Manley-Rowe relations (\ref{M-R}) taking
into account the relation $\omega _{31}\approx \omega _{32}\approx \omega _0$
(one can infer that an error in making such an approximation is of the order
of $\eta =\left( \omega _{31}-\omega _{32}\right) /\omega _{31}=\omega
_T/\omega _{31}\ll 1$). The conservation of the optical intensity may be
written in variables $u_m$ as follows:
\begin{equation}
u_1^2(\zeta )+u_2^2(\zeta )=1.  \label{C1}
\end{equation}
Other ''constants of motion'' following from the Manley-Rowe relations (\ref
{M-R}) are:
\begin{eqnarray}
u_1^2(\zeta )+u_T^2(\zeta ) &=&B,  \label{C2} \\
u_T^2(\zeta )-u_2^2(\zeta ) &=&C,  \label{C3}
\end{eqnarray}
where constants $B$ and $C$ are determined from the boundary conditions at
the $\zeta =0$.

Taking into account the assumed above closed values of the optical dipole
moments: $d_{31}\approx d_{32}$ , one concludes that the quantity $g_0^2$ is
also approximately constant over all the optical path, and it is
approximately equal to $\left( 2\pi d_{31}^2/\hbar ^2c\right) I$. All this
allows one to write the amplitude and the phase Maxwell equations in
variables $u_m,\Phi $ as follows:

\begin{mathletters}
\label{Maxw3}
\begin{eqnarray}
\frac{du_1}{d\zeta } &=&-u_2u_T\sin \Phi , \\
\frac{du_2}{d\zeta } &=&u_1u_T\sin \Phi , \\
\frac{du_T}{d\zeta } &=&u_1u_2\sin \Phi , \\
\frac{d\Phi }{d\zeta } &=&\frac{u_T^2\left( u_2^2-u_1^2\right) -u_1^2u_2^2}{%
u_1u_2u_T}\cos \Phi ,
\end{eqnarray}
where we have assumed that the atomic dipole phases $\vartheta _{ns}$ do not
change along the propagation path. The optical length can now be re-written
as:
\end{mathletters}
\begin{equation}
\zeta =\frac{\pi N}{\hbar ^2c}\mu \left( \frac{8\pi }cI\right) ^{-1/2}\sqrt{%
\omega _T\omega _0}z.  \label{zeta}
\end{equation}

The constants (\ref{C1}), (\ref{C2}) reduce the problem to solve of the set
consisting of the first and the last of Eqs. (\ref{Maxw3}). The right-hand
side of Eq. (\ref{Maxw3} d) can be transformed to the following form by use
of the first three Eqs. (\ref{Maxw3}):
\[
\frac{u_T^2\left( u_2^2-u_1^2\right) -u_1^2u_2^2}{u_1u_2u_T}\cos \Phi =\frac{%
\cos \Phi }{\sin \Phi }\frac 1{u_1u_2u_T}\frac{d\left( u_1u_2u_T\right) }{%
d\zeta }.
\]
Therefore, the equation for the phase can be re-written as
\[
\frac{d\Phi }{d\zeta }=\frac{\cos \Phi }{\sin \Phi }\frac d{d\zeta }\left(
\ln \left( u_1u_2u_T\right) \right) ,
\]
which can immediately be integrated to give the fourth constant of motion:
\begin{equation}
u_1u_2u_T\cos \Phi =\Pi .  \label{const4}
\end{equation}
The value of the constant $\Pi $ can be determined from the known values of $%
u_i$ and $\Phi $ at the entrance to the medium, $\zeta =0$. The expression (%
\ref{const4}) is used then to express $\sin \Phi =\sqrt{1-\cos ^2\Phi }$ in
terms of the conserved quantity $\Pi $, and substitute it, together with the
constants given in Eqs.(\ref{C1}), (\ref{C2}), into the Eq. (\ref{Maxw3} a).
This gives:
\begin{equation}
\zeta =\pm \frac 12%
\displaystyle\int %
\limits_{u_1^2(\zeta =0)}^{u_1^2}\frac{d\left( u_1^2\right) }{\sqrt{%
u_1^2\left( 1-u_1^2\right) \left( B-u_1^2\right) -\Pi ^2}}.  \label{EllF}
\end{equation}
The integral in Eq. (\ref{EllF}) can be reduced to the elliptic integral. In
this paper we consider the case of the THz radiation generation. This
situation corresponds to $u_T(\zeta =0)=0,\,B=u_{10}^2\equiv u_1^2(\zeta
=0),\,\Pi =0$, and to the constant $\cos \Phi $:
\begin{equation}
\cos \Phi (\zeta )=0.  \label{cos0}
\end{equation}
Introduction of a quantity $y=u_1/u_{10}$ leads to the elliptic integral of
the first kind \cite{abr} in standard form:
\[
\zeta =\pm
\displaystyle\int %
\limits_{y_0}^y\frac{dy}{\sqrt{\left( 1-y^2\right) \left(
1-u_{10}^2y^2\right) }}.
\]
The solution can be written then through sinus Jacobi elliptic functions:
\begin{mathletters}
\label{sn}
\begin{eqnarray}
u_1^2(\zeta ) &=&u_{10}^2 sn^2\left[ \left( \zeta +\zeta _0\right) ;u_{10}%
\right] , \\
u_2^2(\zeta ) &=&1-u_{10}^2 sn^2\left[ \left( \zeta +\zeta _0\right) ;u_{10}%
\right] , \\
u_T^2(\zeta ) &=&u_{10}^2\left( 1- sn^2\left[ \left( \zeta +\zeta _0\right)
;u_{10}\right] \right) .
\end{eqnarray}
The solution indicates that as the optical length increases, energy is
periodically transferred between the e.m. waves. The initial condition $%
u_T(\zeta =0)=0$ requires that the constant $\zeta _0$ is equal to $\pm {\bf %
K}(u_{10})$, the complete elliptic integral
\end{mathletters}
\[
{\bf K}(u_{10})=%
\displaystyle\int %
\limits_0^1\frac{dy}{\sqrt{\left( 1-y^2\right) \left( 1-u_{10}^2y^2\right) }}%
,
\]
which is a quarter-period of the function $sn$. Thus, the period of
intensity oscillations is equal to $2{\bf K}(u_{10})$ and the maximum
possible power transferred to the THz radiation: $u_T^2=u_{10}^2$ occurs in
a length
\begin{equation}
\zeta _{max}={\bf K}(u_{10}).  \label{zmax}
\end{equation}
The value of the function ${\bf K}(u_{10})$ is close to $\pi /2$ at $%
u_{10}\ll 1$ and it is increased with $u_{10}$ up to infinity at the limit $%
u_{10}=1$ \cite{abr}. For $u_{20}^2=1-u_{10}^2\ll 1$ a half period can be
approximated by
\begin{equation}
{\bf K}(u_{10})=\frac 12\ln \left( \frac{16}{u_{20}^2}\right)  \label{kln}
\end{equation}

In terms of the real intensity, the Eq. (\ref{sn} c) for the THz radiation
can be written:
\begin{equation}
I_T(\zeta )=I_{31}(\zeta =0)\frac{\omega _T}{\omega _{31}}\left( 1-cd^2\left[
\zeta ;u_{10}\right] \right) .  \label{IT}
\end{equation}
We observe that at the length $\zeta _{max}={\bf K}(u_{10})$ the optical
field is converted into the THz radiation with unity photon efficiency.

The obtained results are confirmed by numerical calculations of the Maxwell
equations (\ref{Maxw1}) with full density matrix calculations without any
approximation. Figure 2 demonstrates the evolution of the e.m. wave
intensities and the relative phase $\Phi $ with the length for real
parameters of the $3^3P_1-\,3^3P_2-4\,^3S_1$ system in $^{24}$Mg atomic
vapor. The optical length in this and all following figures is plotted in
terms of a single-atom absorption cross-section for the optical field $\tau
=\left( 3\pi c^2/2\omega _{_{31}}^2\right) Nz$, which has an advantage of
being independent of total intensity $I$. The length $\zeta $ introduced in
Eq. (\ref{zeta}) is connected to $\tau $ by
\[
\zeta =\tau \frac \mu {d_{31}}\sqrt{\frac{\omega _T}{\omega _{31}}}\left(
\frac{\hbar \gamma _{31}}{d_{31}}\right) \left( \frac{8\pi }cI\right)
^{-1/2}.
\]
In order to compare the numerical and analytical calculations, the results
presented in Fig. 2 are calculated for $\Gamma =0$, $\Delta _{31}=\Delta
_{32}=0$, and vapor temperature $T=10^{-3}K$ (i.e., in fact, for $v_z=0$).
However, the atomic data (wavelengths, dipole moments and decay rates) are
real. In particular, the spontaneous relaxation rates are not equal: $\gamma
_{32}=1.66\gamma _{31}$. The intensities at the entrance to the medium are
chosen so that $u_{20}^2=0.59\cdot 10^{-4}$. The dotted line in Fig. 2(b) is
a calculation with formula (\ref{IT}) for $u_{20}^2=0.55\cdot 10^{-4}$. One
can see that the correspondence not only in the shape but also in
quantitative characteristics of the process is excellent. The difference in
parameter $u_{20}^2$ is obviously due to the difference of spontaneous
relaxation rates. As expected from analytical calculations, the transfer of
energy to the THz radiation occurs with unity photon efficiency (the maximum
value of $I_T$ is exactly equal to $0.0021I_{31}(\zeta =0)$) at the length $%
\tau _{max}=7.6\cdot 10^5$ ($\zeta _{max}=5.87$, this value is quite close
to that calculated from Eq. (\ref{kln}) for $u_{20}^2=0.59\cdot 10^{-4}$: $%
\zeta _{max}=6.25$). Such a perfect frequency conversion is due to the
negligible decay of the dark state given by $\Gamma $ and terms to the
second order in $g_T\,/\,\gamma $. For the parameters of Fig. 2, the maximum
THz intensity $I_T=0.0021I_{31}(\zeta =0)$ corresponds to the Rabi frequency
of only $g_T\approx 8\cdot 10^{-4}\,\gamma _{31}$. The behavior of the phase
$\Phi $ in Fig. 2(c) also follows the law $\cos \Phi (\zeta )=0$ obtained
analytically. The jumps in the phase occur at points where the intensity of
the field being absorbed approaches zero, according to Eqs. (\ref{Maxw1}
(b,d)).

In reality, however, we do have both the Doppler broadening and the
relaxation rate $\Gamma $ of the coherence between states $\left|
1\right\rangle $ and $\left| 2\right\rangle $. Quantitatively, this
considerably modifies the process. In Fig. 3 the spatial dependence of the
field intensities and the phase $\Phi $ are plotted for the vapor
temperature $T=800$ $K$ (this gives the saturated vapor density of $%
N=1.7\cdot 10^{15}\,cm^{-3}$ and corresponds to the most probable velocity
of atoms of $v_p=7.5\cdot 10^4\,cm/sec$), $\Gamma =10^{-4}\gamma _{31}$ (0.6
kHz), input Rabi frequencies $g_{31}\left( \tau =0\right) =60\gamma
_{31},\,\,g_{32}\left( \tau =0\right) =20\gamma _{31}$ (corresponding to
laser intensities of $I_{31}=18.8\,W/cm^2$ and $I_{32}=2.1\,W/cm^2$). We see
that the dynamics of the intensities and the phase does not change
qualitatively as compared to the case of negligible decay of the dark state
(Fig. 2). The THz radiation is generated and reaches its maximum, $%
I_T=39\,mW/cm^2$ (Rabi frequency $g_T\approx 4.6\cdot 10^{-3}\,\gamma _{31}$%
) at the length $\tau =2.2\cdot 10^6$ (for $T=800$ $K$ this corresponds to
the real length of the gas cell of $z=4\,cm$). A minor quantitative
difference compared to the Fig. 2 case is that the optical length scale of
the oscillations increases. This is simply a consequence of the optical
length dependence on the total intensity, Eq. (\ref{zeta}). However, the
Doppler broadening, together with the rate $\Gamma $, considerably influence
the efficiency of the frequency conversion. The efficiency reaches a maximum
when the atoms are prepared in an almost dark state so that the energy
dissipation is very weak. The dark state preparation relies in present case
on the optical pumping. Therefore, the optical pumping rate should be much
larger than the dark state decay in order to allow the population trapping
in $\left| NC\right\rangle $. This requirement results in the following
condition \cite{kors97}:
\[
\frac{g_0^2}{\gamma ^2+\Delta ^2}\gg \frac \Gamma \gamma ,
\]
where detuning $\Delta $ includes the Doppler shift: $\Delta =\Delta
_{31}-k_{31}v_z\approx \Delta _{32}-k_{32}v_z$. For the resonance $\Delta
_{31}=\Delta _{32}=0$ and large Doppler broadening $k_{31}v_p\gg \gamma $,
this condition reduces to
\begin{equation}
g_0^2\gg \frac \Gamma \gamma \left( k_{31}v_p\right) ^2  \label{g0-cond}
\end{equation}
or, in terms of intensity,
\begin{equation}
I\gg \frac \Gamma \gamma \left( \frac{k_{31}v_p}\gamma \right) ^2\frac{16\pi
^2\hbar \gamma c}{3\lambda _{31}^3}.  \label{I-cond}
\end{equation}
For the here considered case of $^{24}$Mg atoms at temperature $800K$, the
total input intensity should be larger than $3.75\cdot 10^3\cdot \Gamma
/\gamma $ [W/cm$^2$]. An importance of the condition (\ref{g0-cond}) is
demonstrated in Fig. 4. While the efficiency is approaching maximum: $\eta
=2.06\cdot 10^{-3}$, for the parameters of Fig. 3 where $\Gamma /\gamma
=10^{-4}$, it drops to $\eta =1.66\cdot 10^{-3}$ for $\Gamma /\gamma =2\cdot
10^{-3}$ in Fig. 4 (a). Moreover, the intensity oscillations decay quite
fast, and at the length $\tau \geq 4\cdot 10^6$ ($\zeta \geq 4.87$) the
optical fields are completely absorbed by the medium (not shown). Increase
of the input intensity in Fig. 4 (b) leads to better trapping, weaker e.m.
energy dissipation and, correspondingly, to larger conversion efficiency
increasing to $\eta =2.00\cdot 10^{-3}$. Of course, with increasing input
intensity the length scale of the conversion increases as well. Parameters
of Fig. 4 (b) correspond to input intensities $I_{31}=470\,W/cm^2$ and $%
I_{32}=52\,W/cm^2$, and to the maximum THz intensity of $I_T=0.94\,W/cm^2$
(Rabi frequency $g_T\approx 2.3\cdot 10^{-2}\,\gamma _{31}$) generated at $%
\tau =1.1\cdot 10^7$ ($z=20\,cm$).

Thus, a practical realization of the proposed here generation scheme
requires large optical length (large atom density), fairly large input
intensities of the optical fields and small decay rate $\Gamma $ of the
atomic coherence $\sigma _{12}$. The rate $\Gamma $ is determined by
uncorrelated laser fluctuations, atomic collisions and other random phase
disturbing processes. It has recently been experimentally demonstrated that,
in a buffer gas, the coherence $\sigma _{12}$ can survive a very large
number of atomic collisions \cite{wyn97,xu98}. For instance, the rate $%
\Gamma <50\,Hz$ has been observed in experiment \cite{wyn97}. The
correlation of the fluctuations in $\omega _{31}$ and $\omega _{32}$
radiations may be achieved if, for example, the $\omega _{32}$ frequency is
generated from $\omega _{31}$ by operation of a Raman laser, as in Ref. \cite
{welleg}.

An important point in nonlinear frequency conversion is a phase matching. In
our scheme, however, the optical frequencies are quite close in magnitude to
each other. Therefore, the wave vectors $k_{3m}$ change approximately
equally due to dispersion of the medium, induced by the interaction with
far-detuned states other than those of the $\Lambda $ system. Thus, this
phase mismatch may be neglected. Moreover, as it has been shown
theoretically and experimentally \cite{jain96}, the phase mismatch can be
compensated for by very small value of two-photon detuning $\left( \Delta
_{31}-\Delta _{32}\right) $, if it does appear.

In the particular case of Mg atoms, considered here as an illustrative
example, the metastable states $3^3P_1$ and $\,3^3P_2$ are not the ground
states. Therefore, one has to prepare an atomic ensemble in metastable
states. This problem may be solved, for example, through a low-energy
high-current electric discharge \cite{god93}. Another problem is that the
state $3^3P_1$ decays spontaneously to the ground state $3^1S_0$. However,
the rate of this decay, $4.3\cdot 10^2\,sec^{-1}=1.2\cdot 10^{-5}\,\gamma
_{31}$ is too slow to produce considerable effect on the process, presented
here. Of course, all these problems may be avoided if one uses atoms with
the metastable states $\left| 1\right\rangle $ and $\left| 2\right\rangle $
being the ground states, like in Sm atoms.

In conclusion, we have proposed a scheme for generation of continuous-wave
THz radiation. Our scheme is based on a nonlinear interaction of three e.m.
waves in atomic media with a $\Lambda $ configuration of levels. An
important ingredient of the scheme is a preparation of atoms in the grey
state. As a consequence, the e.m. fields propagate through the medium with
very weak dissipation of energy. Therefore the process may be considered as
a parametric difference-frequency generation, where the THz frequency is the
difference between two optical ones. We have obtained an analytical solution
of the e.m. propagation problem which demonstrates that the intensities
oscillate along the propagation path in a form of Jacobi elliptic functions.
This allows us to predict that the photon conversion efficiency approaches
unity in this technique, and to estimate the optical length at which the
energy transfer from the optical field into the THz one is maximum. The
analytical solution is confirmed by numerical calculations taking into
account Doppler broadening and relaxation of the grey state. These
calculations show that the efficiency of the THz generation remains still
very high in real situations, if the input optical intensity is sufficiently
high to satisfy the condition (\ref{I-cond}). We note finally that the
tuning of obtained THz radiation may be realized by shifting the metastable
state energies via strong static magnetic or/and electric fields.

\section{Acknowledgments}

We are very grateful to Prof. L. Windholz for his continuous interest to
this work and useful discussions. D.V. Kosachiov thanks the members of the
Institut f\"{u}r Experimentalphysik, TU Graz, for hospitality and support.
This study was supported by the Austrian Science Foundation under project
No. P 12894-PHY.

E. Korsunsky's e-mail address is e.korsunsky@iep.tu-graz.ac.at.

\newpage

{\bf \centerline{\Large{\bf Figure captions}}}

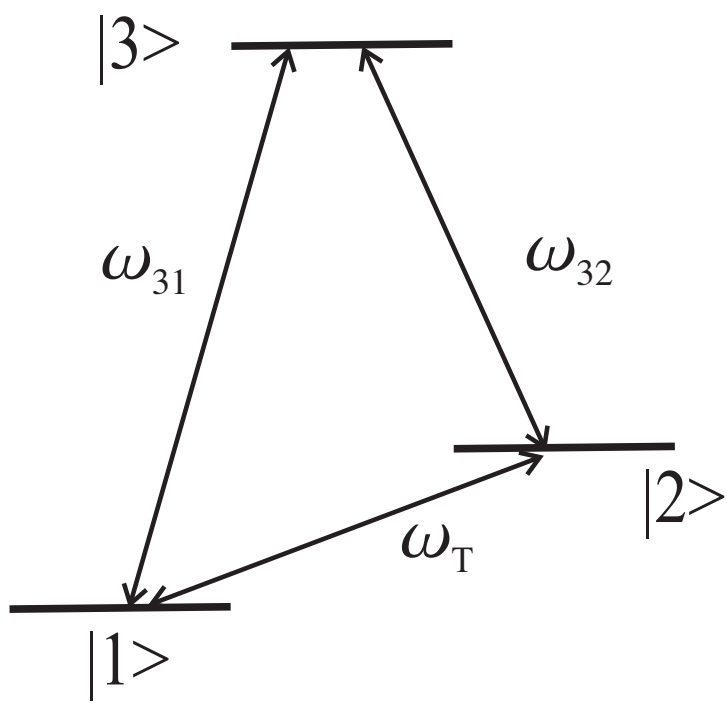
Fig. 1. Closed $\Lambda $ system with two metastable states $\left|
1\right\rangle $ and $\left| 2\right\rangle $. $\omega _{31}$ and $\omega
_{32}$ are the optical frequencies, $\omega _{T}$ is the THz-range frequency.

\medskip

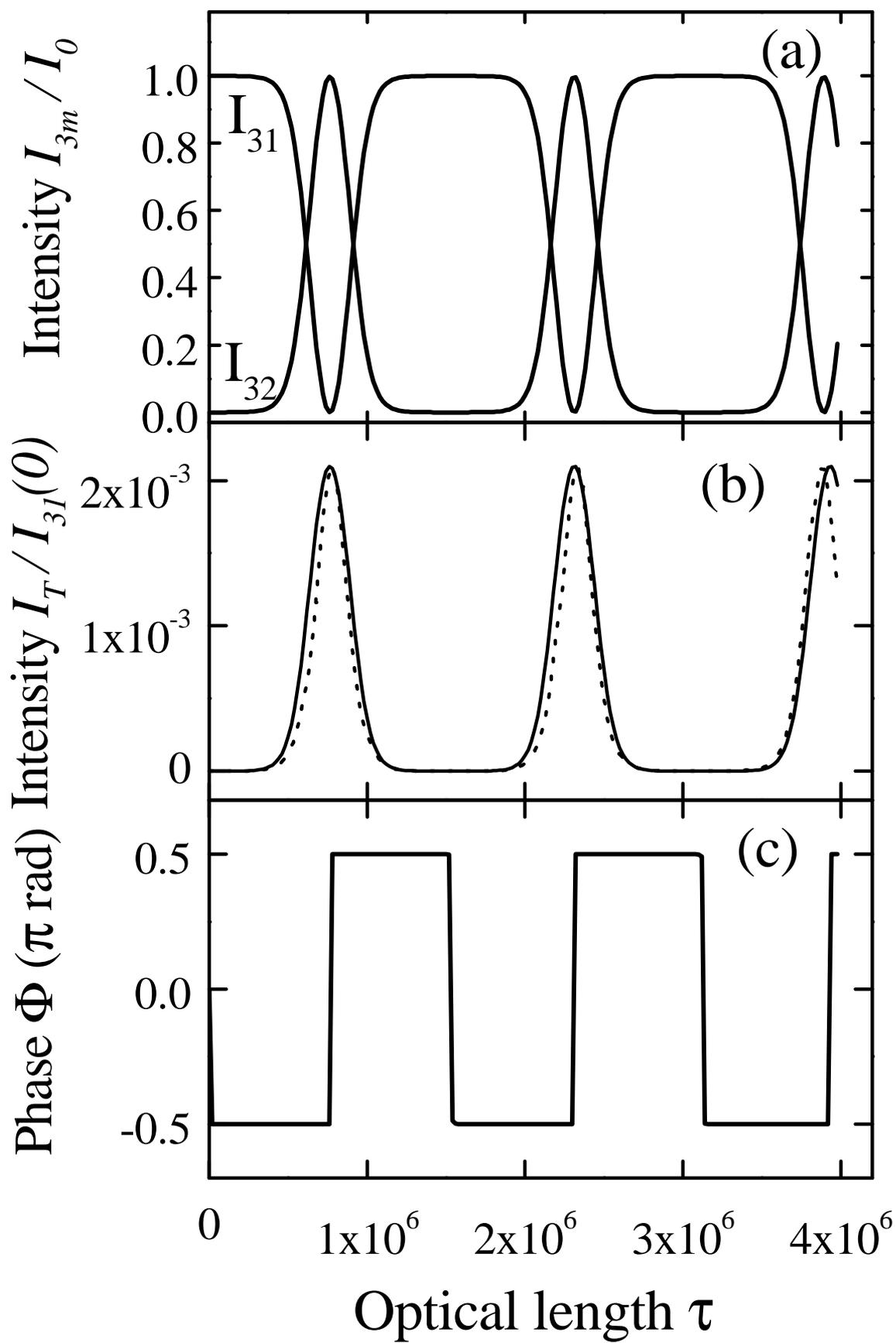
Fig. 2. Spatial variations of the optical (a) and THz (b) field intensities
and the relative phase $\Phi $ (c) in a vapor of $^{24}$Mg atoms interacting
with radiation in a closed $\Lambda $ configuration of levels $%
3^{3}P_{1}-3\,^{3}P_{2}-$ $\,4^{3}S_{1}$. For this system, the relaxation
rates are $\gamma _{31}=3.46\cdot 10^{7}$ sec$^{-1}$, $\gamma
_{32}=1.66\gamma _{31}$, $\gamma _{21}=2.6\cdot 10^{-14}\gamma _{31}$, the
wavelengths $\lambda _{31}=517.27\,nm$, $\lambda _{31}=518.36\,nm$. Other
parameters are: vapor temperature $T=10^{-3}$ $K$, $\Gamma =0$, detunings $%
\Delta _{31}=\Delta _{32}=0$, Rabi frequencies of input fields $g_{31}\left(
\tau =0\right) =10\gamma _{31},\,\,g_{32}\left( \tau =0\right) =0.1\gamma
_{31}$ and $g_{12}\left( \tau =0\right) =0$. The dotted curve in (b) is a
calculation with formula (\ref{IT}) for $u_{20}^{2}=0.55\cdot 10^{-4}$.

\medskip

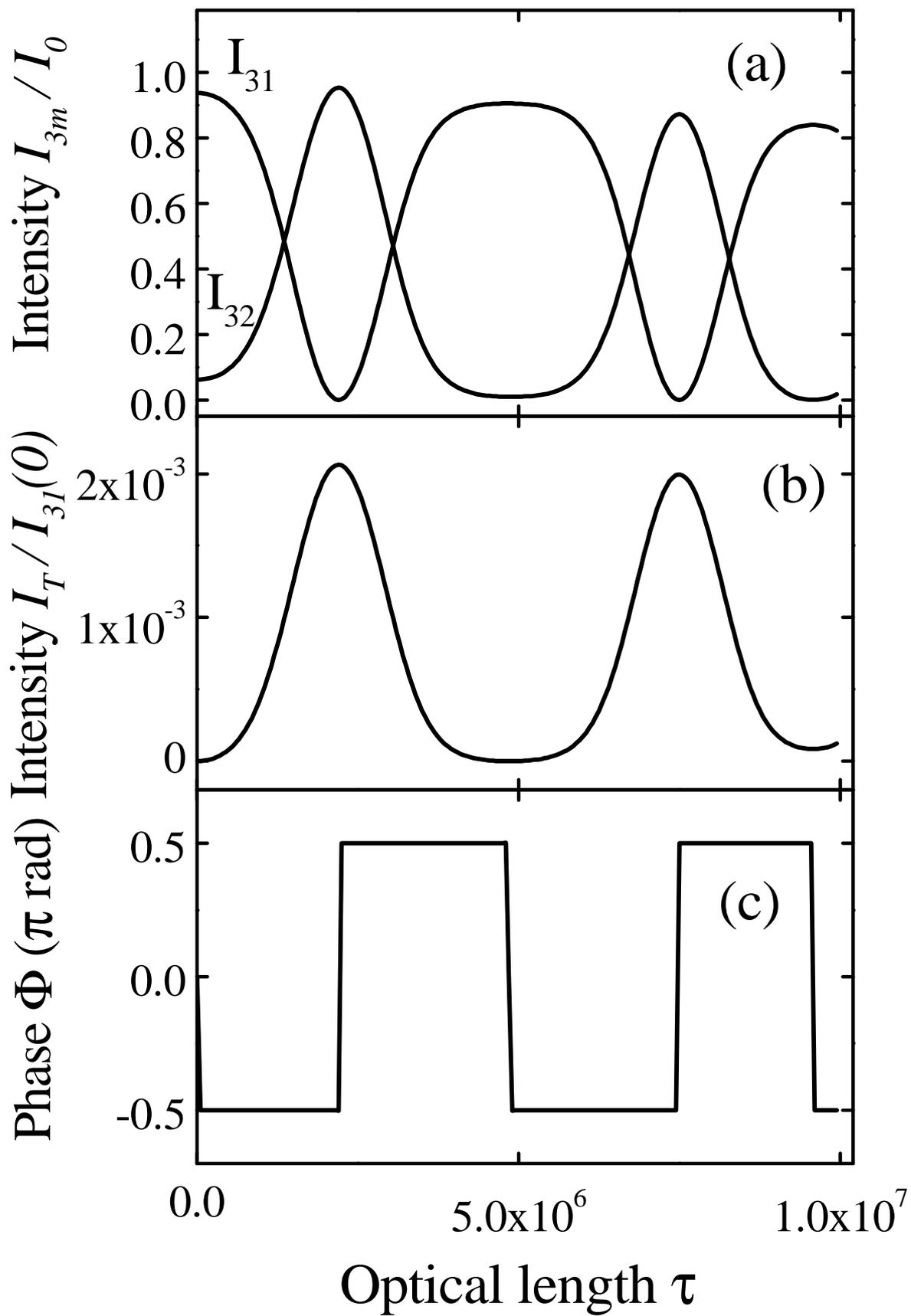
Fig. 3. Spatial variations of the optical (a) and THz (b) field intensities
and the relative phase $\Phi $ (c) in a vapor of $^{24}$Mg atoms for vapor
temperature $T=800$ $K$, $\Gamma =10^{-4}\gamma _{31}$, detunings $\Delta
_{31}=\Delta _{32}=0$, Rabi frequencies of input fields $g_{31}\left( \tau
=0\right) =60\gamma _{31},\,\,g_{32}\left( \tau =0\right) =20\gamma _{31}$
and $g_{12}\left( \tau =0\right) =0$. Other parameters are the same as in
Fig. 2.

\medskip

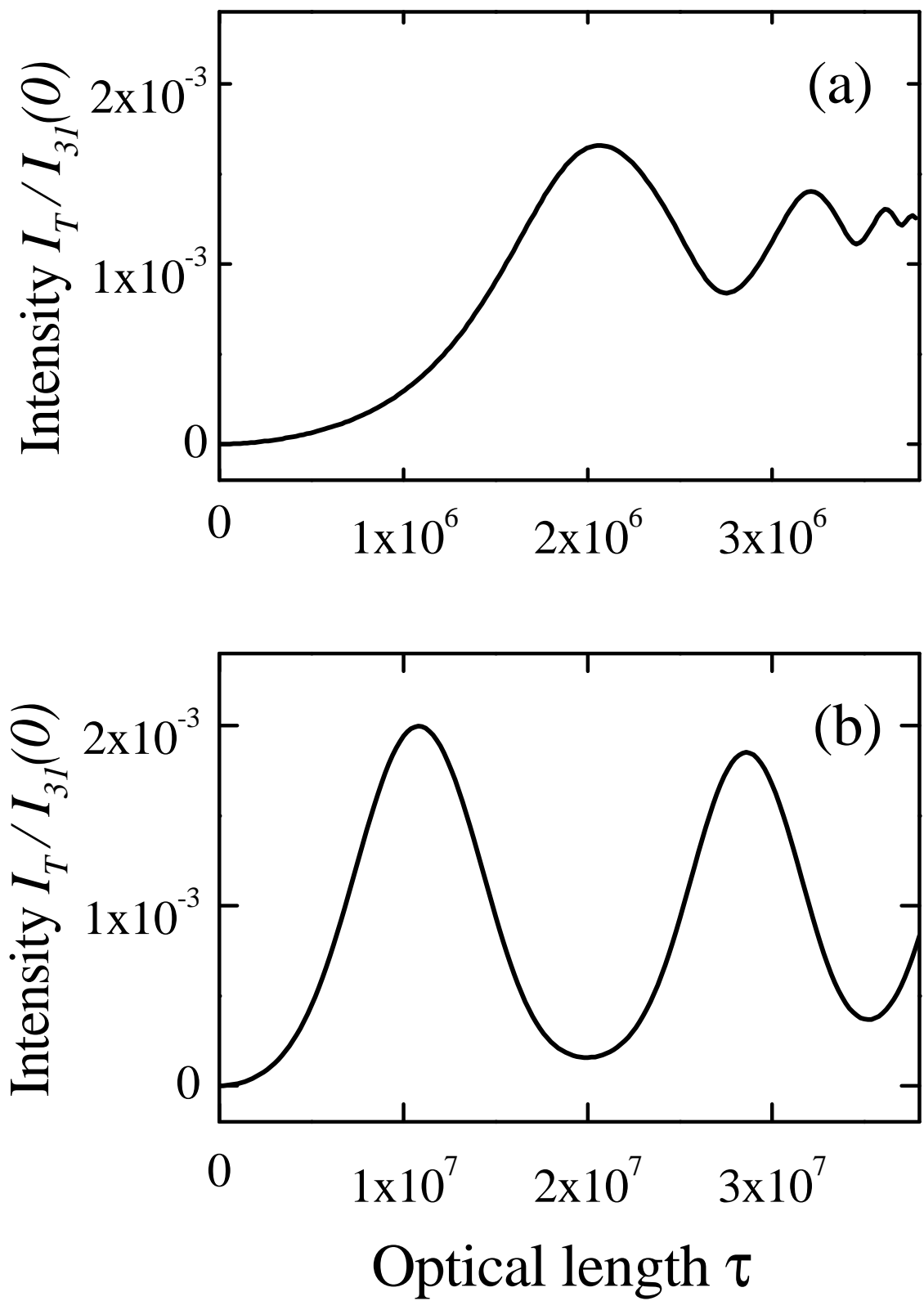
Fig. 4. Spatial variations of the THz radiation intensity in a vapor of $%
^{24}$Mg atoms for $\Gamma =2\cdot 10^{-3}\gamma _{31}$ and Rabi frequencies
of input fields: (a) $g_{31}\left( \tau =0\right) =60\gamma
_{31},\,\,g_{32}\left( \tau =0\right) =20\gamma _{31}$ and (b) $g_{31}\left(
\tau =0\right) =300\gamma _{31},\,\,g_{32}\left( \tau =0\right) =100\gamma
_{31}$. Other parameters are the same as in Fig. 3. Notice the different
length scales in (a) and (b).

\end{document}